\documentclass[twoside,twocolumn,english,prc,amsmath,a4paper,myheadings]{revtex4}
\usepackage[T1]{fontenc}
\usepackage{geometry}
\usepackage{amsmath}
\usepackage{graphicx}
\usepackage{amssymb}

\makeatletter
\def\@oddhead{\rightmark \hfill  Lambda over Kaon Enhancement  \hfill \thepage}
\def\@evenhead{\thepage \hfill K. Werner\hfill}
\def\fnum@table{\tablename~{\bf\thetable}}
\def\fnum@figure{\figurename~{\bf\thefigure}}
\def\tablename{\footnotesize{\bf Table}}
\def\figurename{\footnotesize{\bf Figure}}

\usepackage{dcolumn}

\def\citet{\cite}

\makeatother

\usepackage{babel}

\begin{document}

\title{Lambda over Kaon Enhancement in Heavy Ion Collisions at Several TeV }

\author{{\normalsize K.$\,$Werner}}

\address{SUBATECH, University of Nantes -- IN2P3/CNRS-- EMN, Nantes, France}

\begin{abstract}
We introduced recently a new theoretical scheme which accounts for
hydrodynamically expanding bulk matter, jets, and the interaction
between the two. Important for the particle production at intermediate
values of transverse momentum ($p_{t}$) are jet-hadrons produced
inside the fluid. They pick up quarks and antiquarks (or diquarks)
from the thermal matter rather than creating them via the Schwinger
mechanism -- the usual mechanism of hadron production from string
fragmentation. These hadrons carry plasma properties (flavor, flow),
but also the large momentum of the transversely moving string segment
connecting quark and antiquark (or diquark). They therefore show up
at quite large values of $p_{t}$, not polluted by soft particle production.
We will show that this mechanism leads to a pronounced peak in the
lambda / kaon ratio at intermediate $p_{t}$. The effect increases
substantially with centrality, which reflects the increasing transverse
size with centrality. 
\end{abstract}
\maketitle
Heavy ion collisions at relativistic energies are expected to lead
to the formation of a quark gluon plasma, which strongly interacts
and behaves as a fluid \citet{intro1,intro2,intro3,intro4}. Nevertheless
it is difficult to directly observe the plasma properties, since the
fluid hadronizes and the corresponding hadrons still interact among
themselves before being detected. It is therefore desirable to find
observables which keep information about the partonic system despite
the hadronization procedure. In this paper we will discuss in which
sense the transverse momentum dependence of the lambda over kaon ratio
is such an observable.

As already observed earlier in AuAu scattering at 200 GeV \citet{star-lda},
also in PbPb collisions at 2.76 TeV there is an impressive increase
of the lambda yield compared to kaons, more and more pronounced with
increasing centrality, as shown by the ALICE collaboration \citet{ali3-lda}.
This phenomenon concerns transverse momenta ($p_{t}$) in the range
between 2 and 6 GeV/c. This so-called {}``intermediate $p_{t}$ range''
is the domain of coalescence models \citet{coa1,coa2,coa3,coa4,coa5},
where hadrons are produced by recombining quarks from the plasma,
to be distinguished from {}``fragmentation'' of partons. However,
a detailed quantitative understanding of the intermediate $p_{t}$
region is still missing, and as discussed in \citet{jetbulk}, it
seems impossible to really separate an {}``intermediate $p_{t}$
region'' from the low and high transverse momentum domain.

In ref. \citet{jetbulk}, we introduced a new theoretical scheme which
accounts for hydrodynamically expanding bulk matter, jets, and the
interaction between the two. The whole transverse momentum range is
covered, from very low to very high $p_{t}$. In \citet{jetbulk},
we show that the new approach can accommodate spectra of jets with
$p_{t}$ up to 200 GeV/c in $pp$ scattering at 7 TeV, as well as
particle yields and harmonic flows with $p_{t}$ between 0 and 20
GeV/c in PbPb collisions at 2.76 TeV. Since out aim is a single model
which is able to cover all phenomena, we will apply the approach of
ref. \citet{jetbulk}, with exactly the same parameters (EPOS2.17v3),
to study lambda and kaon production, and try to understand the {}``lambda
over K peak''.

Let us briefly recall the essential features of the new approach,
which are relevant for the discussion of this paper. All the details
can be found in \citet{jetbulk}. The basis are multiple scatterings
(even for $pp$), where a single scattering is a hard elementary scattering
plus initial state radiation, the whole object being referred to as
parton ladder. The corresponding final state partonic system amounts
to (usually two) color flux tubes, being mainly longitudinal, with
transversely moving pieces carrying the $p_{t}$ of the partons from
hard scatterings. These flux tubes constitute eventually both bulk
matter (which thermalizes, flows, and finally hadronizes) and jets,
according to some criteria based on partonic energy loss. 

\begin{figure}[tb]
\begin{centering}
\includegraphics[scale=0.15]{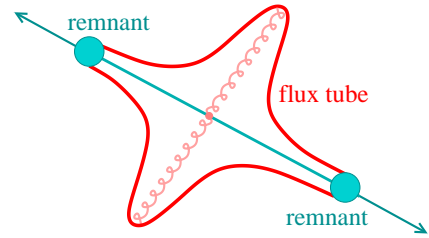}
\par\end{centering}

\caption{(Color online) A single hard scattering, leading to two flux tubes
with transversely moving parts (kinky strings). \protect \\
\label{stgfr1}}

\end{figure}

\begin{figure}[tb]
\begin{centering}
\includegraphics[scale=0.15]{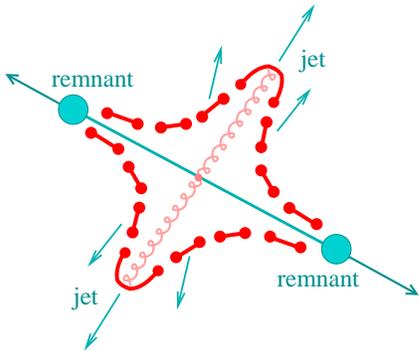}
\par\end{centering}

\caption{(Color online) Flux tube breaking via $q-\bar{q}$ production, which
screens the color field (Schwinger mechanism).\label{stgfr2}}

\end{figure}

Let us take the simple case of a single hard scattering producing
two gluons, without initial state radiation. This leads to two flux
tubes, with one transversely moving piece each, corresponding to the
hard gluons, as shown in fig. \ref{stgfr1}. %
These flux tubes do not only cover central rapidities, since they
stretch from projectile to target remnant. Including initial state
radiation (which is automatically included in all calculations), will
add more transversely moving pieces, leading to a complicated three-dimensional
dynamics. But despite the complicated details, the flux tubes remains
essentially longitudinal, with some transversely moving parts (kinks
in the string language). The Flux tubes (strings) will expand and
at some stage break via the production of quark-antiquark or diquark-antidiquark
pairs, as seen in fig. \ref{stgfr2}. %
The string segments are identified with hadrons. Those close to the
transversely moving pieces carry the large momentum coming from the
partons of the hard scattering -- they constitute the jets, indicated
by the arrows in fig. \ref{stgfr2}.

In heavy ion collisions and also in high multiplicity events in proton-proton
scattering at very high energies, there are many elementary scatterings,
and therefore many flux tubes. Their density will be so high that
they cannot decay independently as described above. Here we have to
modify the procedure as discussed in the following. The starting point
are still the flux tubes (kinky strings) originating from elementary
collisions, as discussed above. These flux tubes finally constitute
both, bulk matter which thermalizes and expands collectively, and
jets. The criterion which decides whether a string piece ends up as
bulk or jet, is based on energy loss. In the following, we consider
a flux tube in matter, where {}``matter'' first means the presence
of a high density of other flux tubes, which then thermalize. 

Three possibilities occur: (A) String segments which have not sufficient
energy to escaspe will constitute matter, they loose their character
as individual strings. This matter will evolve hydrodynamically and
finally hadronize ({}``soft hadrons''). (B) String segments having
sufficient energy to escape and being formed outside the matter, constitute
jets ({}``jet-hadrons''). (C) There are finally also string segments
produced inside matter or at the surface, but having enough energy
to escape and show up as jets ({}``jet-hadrons''). They are affected
by the flowing matter ({}``fluid-jet interaction''). 

The criterion for a string segment to leave the matter or not, is
based on energy loss, as discussed in\citet{jetbulk}. In case (B),
high energy flux tube segments will leave the fluid, providing jet-hadrons
via the usual Schwinger mechanism of flux-tube breaking caused by
quark-antiquark or diquark-antidiquark production. 

\begin{figure}[tb]
\begin{raggedright}
\vspace*{0.5cm}
\par\end{raggedright}

\begin{raggedright}
{\large \hspace*{1.2cm}(a)}
\par\end{raggedright}{\large \par}

\vspace*{-0.5cm}

\begin{centering}
\includegraphics[scale=0.15]{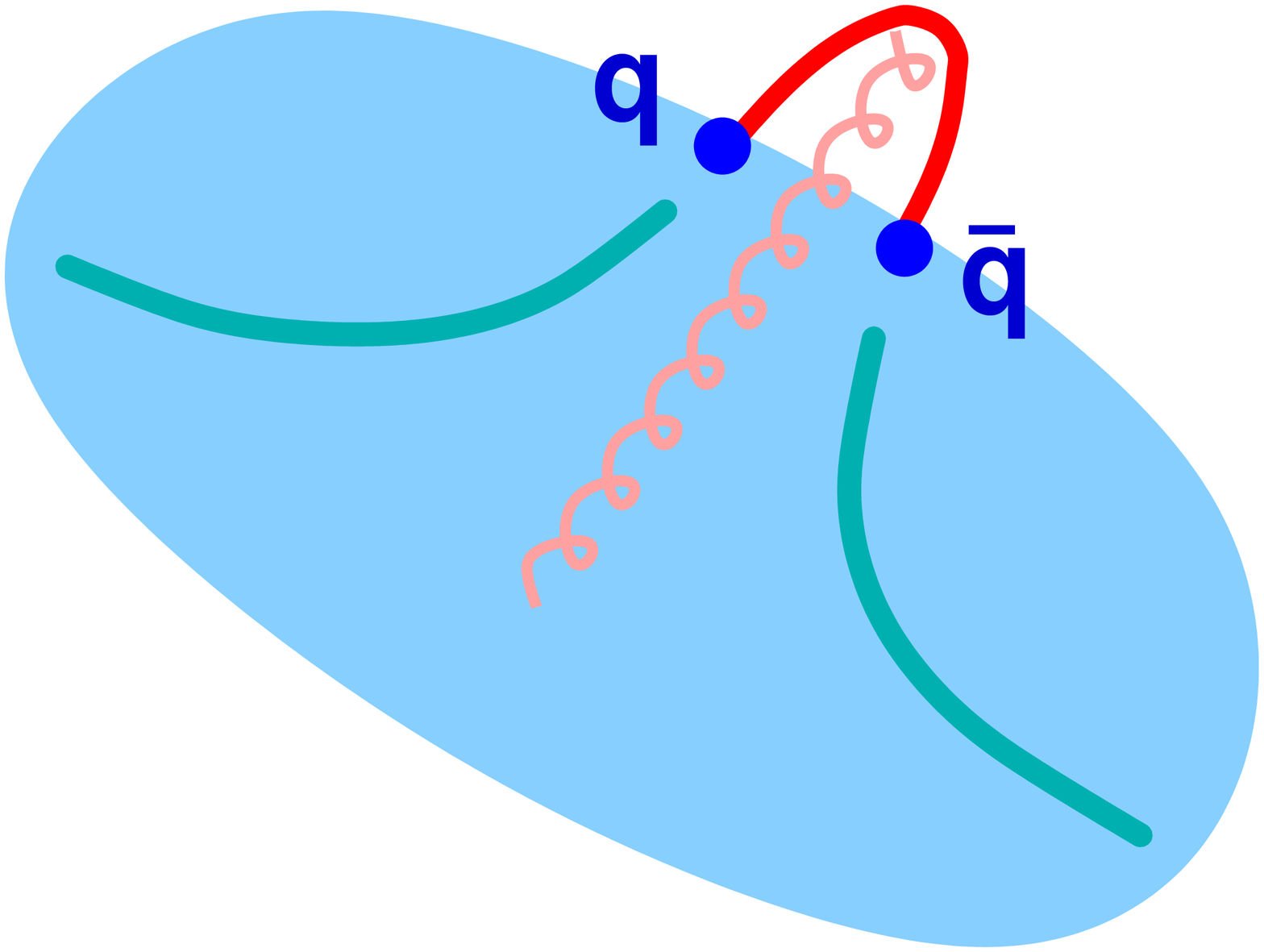}
\par\end{centering}

\begin{raggedright}
{\large \hspace*{1.2cm}(b)}
\par\end{raggedright}{\large \par}

\vspace*{-0.5cm}

\begin{centering}
\includegraphics[scale=0.15]{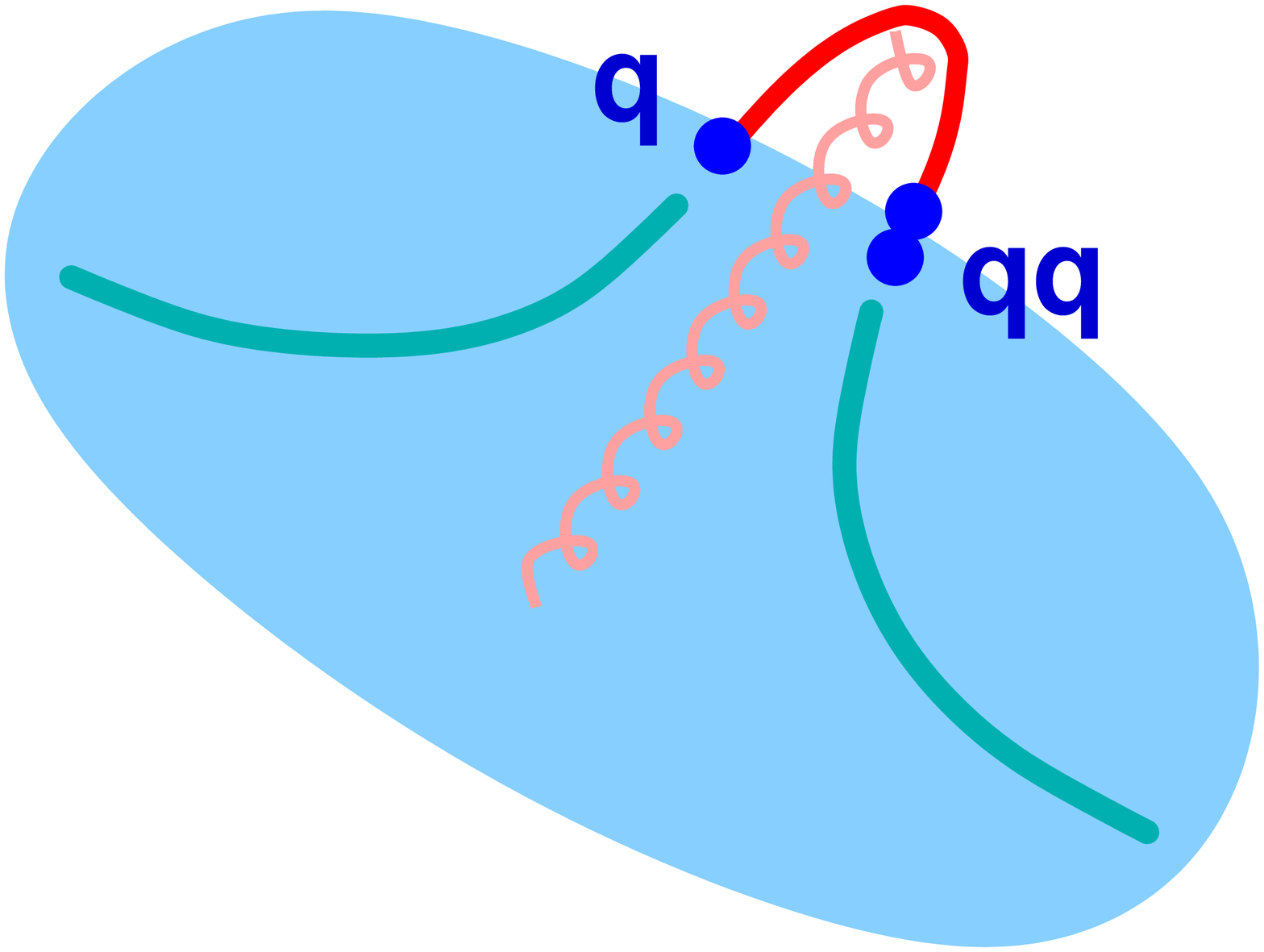}
\par\end{centering}

\caption{(Color online) Escaping string segment, getting it's endpoint partons
from the fluid. We show the case of a quark and an antiquark (a) and
of a quark and a diquark (b). The rest of the string dissolves in
matter.\label{stgfr3}}

\end{figure}

Interesting is case (C). The jet-hadrons are produced still inside
matter or at the surface, but they escape. Here we assume that the
quark, antiquark, diquark, or antidiquark needed for the flux tube
breaking is provided by the fluid with properties (momentum, flavor)
determined by the fluid rather than the Schwinger mechanism, whereas
the rest of the string dissolves in matter, see fig. \ref{stgfr3}.
Considering transverse fluid velocities up to 0.7c, and thermal parton
momentum distributions, one may get a {}``push'' of a couple of
GeV to be added to the transverse momentum of the string segment.
This will be a crucial effect for intermediate $p_{t}$ jet-hadrons
and explains azimuthal asymmetries up to quite large values of $p_{t}$,
as discussed in very detail in \citet{jetbulk}. 

Even more important for the present discussion are two other effects:
The quark (antiquark) flavors are determined from Bose-Einstein statistics,
with more strangeness production compared to the Schwinger mechanism.
And the probability $p_{\mathrm{diq}}$ to have a diquark rather than
an antiquark will be bigger compared to a highly suppressed diquark-antidiquark
breakup in the Schwinger picture ($p_{\mathrm{diq}}$ is a parameter).

\begin{figure}[tb]
\begin{raggedright}
{\large \hspace*{0.8cm}(a)}
\par\end{raggedright}{\large \par}

\vspace*{-0.9cm}

\begin{centering}
\includegraphics[angle=270,scale=0.32]{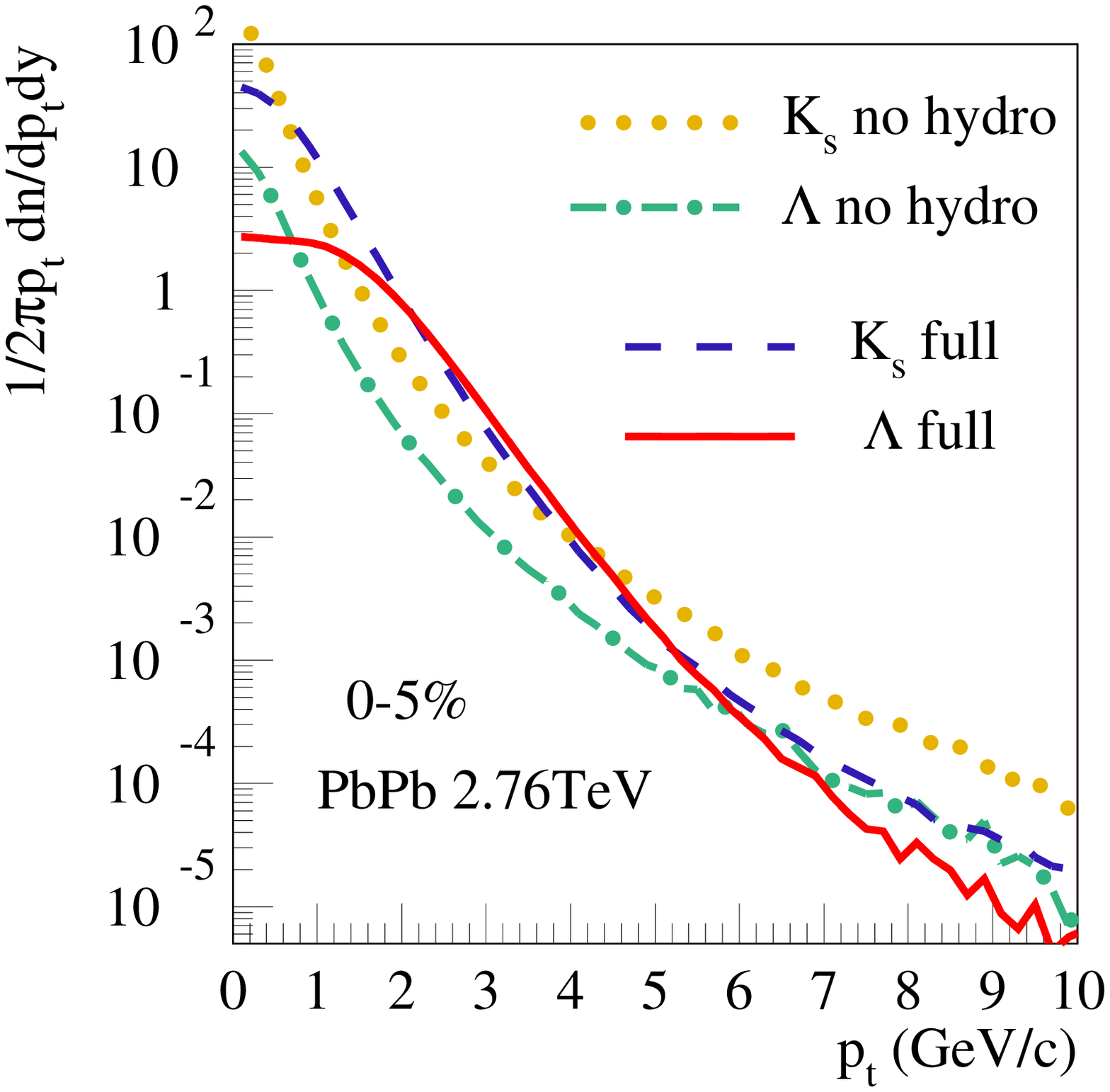}
\par\end{centering}

\begin{raggedright}
\vspace*{-0.2cm}
\par\end{raggedright}

\begin{raggedright}
{\large \hspace*{0.8cm}(b)}
\par\end{raggedright}{\large \par}

\vspace*{-0.9cm}

\begin{centering}
\includegraphics[angle=270,scale=0.32]{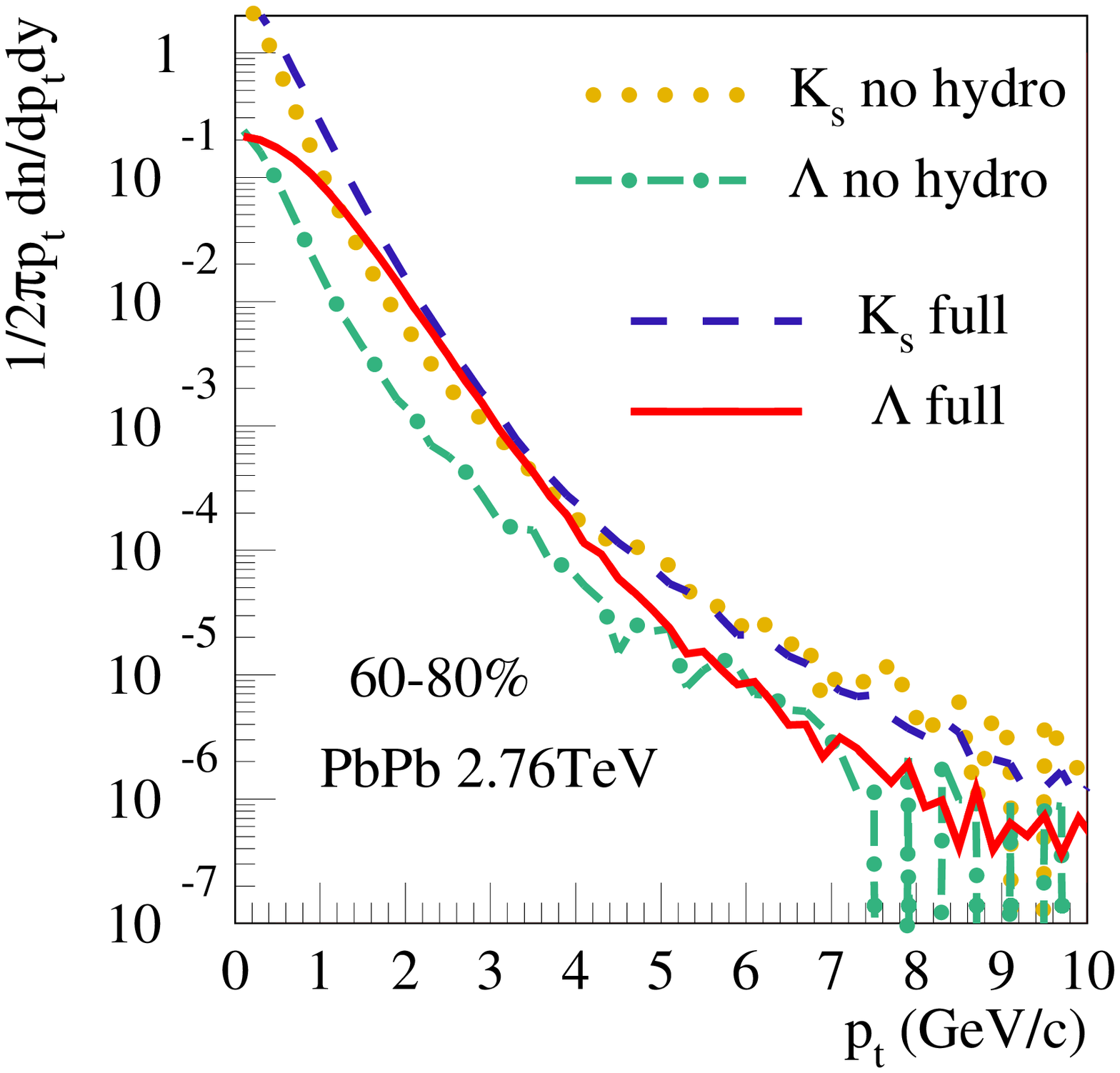}
\par\end{centering}

\caption{(Color online) Transverse momentum spectra for lambdas and kaons ($K_{s}$)
for central rapidities in central (a) and peripheral (b) PbPb collisions
at 2.76 TeV. We show the full model calculations, and also the ones
without hydro evolution.\label{fig:ldaka}}

\end{figure}

All these effects are important concerning lambda and kaon production.
In fig. \ref{fig:ldaka}, %
we show transverse momentum spectra for lambdas and kaons for central
rapidities in central and peripheral PbPb collisions at 2.76 TeV.
The spectra for the calculations without hydrodynamic evolution (pure
string decay) are quite similar for lambdas and kaons, they essentially
differ by a factor, simply due to the fact that the relative probability
of diquark-antidiquark to quark-antiquark breakup is small. However,
the situation is completely different, in case of the full calculation.
Both lambdas and kaons increase at intermediate $p_{t}$, but much
more in case of lambdas. This is first of all due to flow, which pushes
the heavier lambdas more than the kaons. 

The effect is magnified due to jet-hadrons carrying fluid properties
(process (C)): There is an additional momentum push from the fluid-jet
interaction, which favors lambdas over kaons, due to the higher number
of quarks of the former ones. Also the yield of lambdas is increased
compared to kaons, because diquarks compared to quarks are less suppressed
when taken from the fluid as compared to the Schwinger mechanism.

\begin{figure}[tb]
\begin{centering}
\includegraphics[angle=270,scale=0.32]{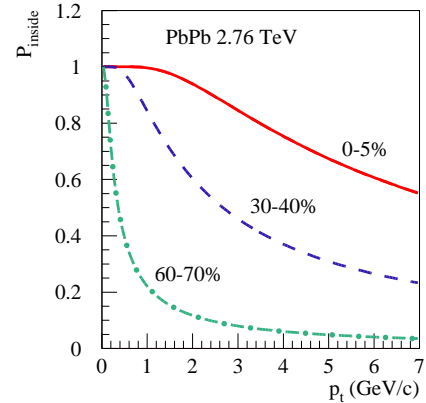}
\par\end{centering}

\caption{(Color online) The estimate $P_{\mathrm{inside}}$ of the probability
to form (pre)hadrons inside the fluid (in the direction parallel to
the impact parameter), as a function of $p_{t}$, for PbPb collisions
at 2.76 TeV. We show the curves for the 0-5\%, the 30-40\%, and the
60-70\% most central events.\label{fig:estim}}

\end{figure}

The effects are similar in central and peripheral PbPb collisions,
but the preferred lambda production compared to kaons is more pronounced
in the central compared to the peripheral events: the lambda curve
crosses the kaon one in case of central, but not in case of peripheral
collisions. The reason is that the number of jet-hadrons carrying
fluid properties depends on their formation times: these hadrons must
have been formed inside the fluid. In fig. \ref{fig:estim}, %
we plot the estimate $P_{\mathrm{inside}}$ of the probability to
form (pre)hadrons inside the fluid, as a function of $p_{t}$, for
different centralities (see \citet{jetbulk}). This probability decreases
strongly towards peripheral collisions, because the transverse sizes
get much smaller. So we get less lambda enhancement at more peripheral
collisions, and the enhancement is also shifted to smaller $p_{t}$.

\begin{figure*}[tb]
\begin{centering}
\includegraphics[angle=270,scale=0.43]{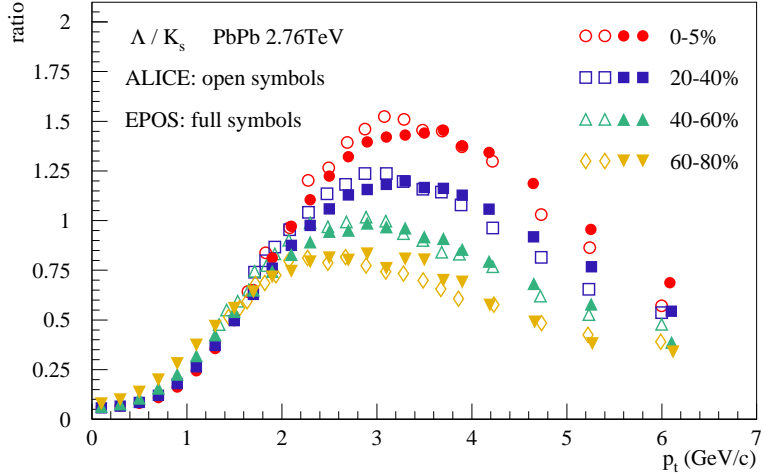}
\par\end{centering}

\caption{(Color online) Lambda over kaon ratio as a function of $p_{t}$ in
PbPb collisions at different centralities. We show our theoretical
results (full symbols) and data from ALICE \citet{ali3-lda} (open
symbols). \label{fig:ratio}}

\end{figure*}

In fig. \ref{fig:ratio}, %
we show the lambda over kaon ratio as a function of $p_{t}$ in PbPb
collisions at different centralities. Our calculations (full symbols)
follow quite well the trend seen in the data (open symbols): going
from peripheral towards central collisions, one observes a more and
more pronounced peak, which also moves to higher $p_{t}$. The calculations
would fit the data even better with a slightly smaller formation time,
but we prefer to use the parameters of ref. \citet{jetbulk}, which
give quite good agreement with many other data sets.

To summarize: We introduced a new theoretical scheme which accounts
for hydrodynamically expanding bulk matter, jets, and the interaction
between the two. This approach covers the whole transverse momentum
range, from very low to very high $p_{t}$. In this framework, we
can reproduce the experimentally observed strong increase of the lambda
over kaon ratio at intermediate values of $p_{t}$. We understand
this effect to be due to a communication between the fluid and jet-hadrons:
these hadrons are composed of a high $p_{t}$ string segment (from
the hard process) and (di)quarks from the fluid, carrying fluid properties.
So the final hadrons are observed at relatively high $p_{t}$, but
nevertheless providing information about the fluid, whereas the soft
hadrons from fluid freeze out carry only small $p_{t}$. The reason
for the strong centrality dependence is the fact that the number of
jet-hadrons having suffered a fluid-jet interaction depends on the
volume: the probability that such a jet-hadron is produced inside
the fluid is more likely for big volumes compared to small ones. The
fact that the enhancement disappears for very peripheral collisions
does not mean that there is no fluid. It means that the volume is
too small for this particular effect to be seen.

\end{document}